\documentclass[11pt]{article}
\usepackage{graphicx}
\usepackage{amsfonts}
  \def\fH{{\cal H}} 
    
\def\fD{{\cal D}}

\usepackage{theorem}
\theorembodyfont{\rmfamily}
\newtheorem{Lem}{Lemma}[section]
\newtheorem{Def}[Lem]{Definition}

\newtheorem{Prop}[Lem]{Proposition}

\newtheorem{Ex}[Lem]{Example}

\newcommand{\qed}{\hbox{\rule{6pt}{6pt}}}

\begin{document}
\title{A note on a parametrically extended entanglement-measure due to Tsallis relative entropy}
\author{Shigeru Furuichi$^1$\footnote{E-mail:furuichi@chs.nihon-u.ac.jp}  \\
$^1${\small Department of Computer Science and System Analysis,}\\
{\small College of Humanities and Sciences, Nihon University,}\\
{\small 3-25-40, Sakurajyousui, Setagaya-ku, Tokyo, 156-8550, Japan}}
\date{}
\maketitle

{\bf Abstract.} 
In the previous paper \cite{FYK}, we mainly studied the mathematical properties of the Tsallis relative entropy for density operators.
As an application of its properties, we adopt a parametrically extended entanglement-measure due to the Tsallis relative entropy in order to measure 
the degree of entanglement. Then the relation between our measure and the relative entropy of entanglement is studied.
\vspace{3mm}

\vspace{3mm}

%\section*{Extended abstract}

\section{Introduction}

Recently, information theory has been rapidly in a progress as quantum information theory \cite{NC}
with the help of the mathematical studies such as operator theory \cite{Furuta} and quantum entropy theory \cite{OP}. 
As one of famous applications of quantum entropy in quantum information theory, 
the quantification of the degree of entanglement is known \cite{Ben,Ved}. To quantify the degree of entanglement,
the von Neumann entropy was firstly used. For pure entangled states, the von Neumann entropy is suitable to quantify the
degree of entanglement, since it does not depend on the choice how to take the partial trace of the subsystems. However, for mixed entangled states,
the von Neumann entropy does not uniquely determine the degree of entanglement. Thus C.H.Bennet et. al. \cite{Ben} introduced the 
entanglement of formation, which is the minimum of the average of von Neuamann entropy for the reduced states, as the degree of
entanglement for the mixed entangled states. Later, V.Vedral et.al. \cite{Ved,Ved2} introduced the relative entropy   
of entanglement, which is the minimum of the relative entropy between the mixed entangled state and the separable (disentangled) state. 
Then they gave the conditions that the entanglement-measure should satisfy. However these quantities have a disadvantage
that the actual calculations of them are difficult. Therefore we proposed the degree of entanglement due to the mutual entropy
which can be regarded as a special case of the entanglement of relative entropy, and then applied it to analyze the Jaynes-Cummings model in \cite{FO,FM,FN}. 
Our entanglement-measure contains both quantum and classical entanglement so that it is not suitable to analyze the purely quantum entanglement itself. 
However it is not difficult to calculate our entanglement-measure due to the mutual entropy and it is enough to estimate a rough degree of entanglement
for the actual models. 

On the other hand, in statistical physics, the Tsallis entropy was defined in \cite{Tsa} by 
$$S_q(X) \equiv -\sum_{x} p(x)^q \ln_q p(x)$$ 
with one parameter $q$ as an extension of Shannon entropy, where $q$-logarithm is defined by $\ln_q(x) \equiv \frac{x^{1-q}-1}{1-q}$
for any nonnegative real number $q$ and $x$, and $p(x) \equiv p(X=x)$ is the probability distribution of the given randam variable $X$.
The Tsallis entropy $S_q(X)$ converges to the Shannon entropy $-\sum_{x} p(x) \log p(x)$ as $q \to 1$, 
since $q$-logarithm uniformly converges to natural logarithm as $q \to 1$. 
The Tsallis entropy plays an essential role in nonextensive statistics, which is often called Tsallis statistics, so that
many important results have been published from the various points of view \cite{AO}.
It is important to study a new entropic quantities, thus we mathematically studied the Tsallis relative entropy in both classical and quantum systems in \cite{FYK}.
In our previous paper, the properties (nonnegativity, unitary invariance and monotonicity) of the Tsallis relative entropy in quantum system has been shown.
In this short paper, we adopt an entanglement-measure due to the Tsallis relative entropy as a special version of the generalized Kulback-Leibler measure of quantum entanglement 
originally introduced in \cite{AR} and then study its properties and give the relation to the relative entropy of entanglement.

\section{Tsallis relative entropy}
After the birth of the Tsallis entropy, recently the uniqueness theorem for the Tsallis entropy was proved in \cite{Suy} 
by introducing the generalized Shannon-Khinchin's axiom.
We introduced the generalized Faddeev's axiom and then modified the uniqueness theorem for the Tsallis entropy 
in \cite{Furu} in the sense that the Faddeev's axiom is simpler than the Shannon-Khinchin's one.
We also characterized the Tsallis relative entropy by introducing the generalized Hobson's axiom in \cite{Furu}. 
These results motivates us to study the Tsallis relative entropy
from the information-theoretical point of view. In this section, we briefly review the mathematical properties of the Tsallis relative entropy in quantum system.
As a noncommutative extention, the quantum Tsallis relative entropy was defined by the following, 
(See \cite{FYK,Ab2,Ab3} and also Chapter II of \cite{AO}.) 
\begin{equation}\label{qqre}
D_q(\rho\vert \sigma) \equiv \frac{Tr[\rho - \rho^q\sigma^{1-q}]}{1-q} 
\end{equation}
for two density operators $\rho$ and $\sigma$ and $0 \leq q < 1$, as one parameter extension of the definition of quantum relative entropy introduced by H.Umegaki \cite{Um}
\begin{equation}
\lim_{q \to 1} D_q(\rho \vert \sigma) = U(\rho\vert \sigma) \equiv Tr[\rho(\log \rho - \log \sigma)].
\end{equation}

For the quantum Tsallis relative entropy $D_q(\rho\vert \sigma)$,
(i) pseudoadditivity and (ii) nonnegativity were shown in \cite{Ab2}, moreover (iii) joint convexity and (iv) monotonicity for projective mesurements,
were shown in \cite{Ab3}.
In addition, we showed in \cite{FYK} the (v) unitary invariance of $D_q(\rho\vert \sigma)$ and 
(vi) the monotonicity of that for the trace-preserving completely positive linear map, without assumption that density operators are invertible.

\begin{Prop} {\bf (\cite{Petz,OP,FYK})}   \label{prop1}
For $0 \leq q < 1$ and any density operators $\rho$ and $\sigma$, the quantum relative entropy $D_q(\rho \vert \sigma)$ has the following properties.
\begin{itemize}
\item[(1)] (Nonnegativity) $ D_q(\rho \vert \sigma) \geq 0 .$
\item[(2)] (Unitary invariance) The quantum Tsallis relative entropy is invariant under the unitary transformation $U$ :
$$D_q(U\rho U^*\vert U\sigma U^*) = D_q(\rho\vert \sigma). $$
\item[(3)]  (Monotonicity) For any trace-preserving completely positive linear map $\Phi$, any density operators $\rho$ and $\sigma$ and $0\leq q <1$, we have
$$ D_q(\Phi(\rho) \vert \Phi(\sigma)) \leq D_q(\rho \vert \sigma). $$
\end{itemize}
\end{Prop}

\section{Degree of entanglement due to Tsallis relative entropy}

Since a separable state is considered that it dose not have any quantum correlation (entanglement), non-separable states is called entangled states. 
We give the definition of a separable (disentangled) state and an entangled state in the following \cite{RW,Pere,Horo3,Horo2}.
\begin{Def}
A state $\kappa$ acting on the composite system $\fH_1 \otimes \fH_2$ is called a separable (disentangled) state if it is represented by
$$\kappa = \sum_i p_i \kappa_1^i \otimes \kappa_2^i, \,\, p_i \geq 0,\,\, \sum_i p_i =1, $$
for states $\kappa_1^i$ and $\kappa_2^i$ acting on the subsystems $\fH_1$ and $\fH_2$, respectively.
It is also called an entangled state if a state is not a separable state.
\end{Def}

The concept of entanglement has been important in quantum information theory, especially quantum teleportation and quantum computing and so on.
Therefore it is important to quantify the degree of entanglement, in order to scientifically treat the concept of entanglement. When we quantify the entanglement, we should
pay the attention whether the considering entangled state is pure or mixed. For pure entangled states, it is easily calculated by 
the von Neumann entropy of the reduced states. We suppose that the entangled states are presented by the density operator $\sigma$ on the tensor Hilbert space $\fH_1 \otimes \fH_2$.
We also present two reduced states by $\sigma_1$ and $\sigma_2$, respectively. Then we have the following triangle inequality \cite{AL}:
 $$ \vert S(\sigma_1)-S(\sigma_2) \vert \leq S(\sigma) \leq S(\sigma_1) + S(\sigma_2). $$
Thus we have $S(\sigma_1) = S(\sigma_2)$ for pure entangled states $\sigma$ on $\fH_1 \otimes \fH_2$, so two von Neumann entropies are equal to each other, which means 
that it does not depend on the choice how to take the partial trace on Hilbert space $\fH_1$ or $\fH_2$. Then the degree of entanglement for the pure entangled states is defined by
$$ E(\sigma) \equiv -Tr[\sigma_1 \log \sigma_1] = -Tr[\sigma_2 \log \sigma_2].$$
However, for mixed entangled states, the degree of entanglement is not uniquely determined by von Nuemann entropy in general.
Then C.H.Bennet et.al. introduced the entanglement of formation for mixed entangled states in the following.
\begin{Def} {\bf (\cite{Ben})}
For mixed entangled states $\sigma = \sum_i p_i \sigma^{(i)}$, where $\sum_i p_i =1$, $p_i \geq 0$ and $\sigma^{(i)} = \vert \phi_i \rangle \langle \phi_i \vert$ are pure entangled states on $\fH_1 \otimes \fH_2$,  
the entanglement of formation is defined by
$$E^F(\sigma) \equiv \min \sum_i p_i S(\sigma_1^{(i)})$$
as a minimun of the average of the von Neumann entropy $S(\sigma_1^{(i)})$ of the reduced states $\sigma_1^{(i)}$ for the pure entangled states $\sigma^{(i)}$, 
where the minimum is taken over all the possible states
$\sigma = \sum_i p_i \sigma^{(i)}$ with $\sigma_1^{(i)} = tr_{\fH_2}\sigma^{(i)}$.
\end{Def} 
There are several measures for the degree of entanglement other than the above. As a famous and an important measure, 
the relative entropy of entanglement introduced in \cite{Ved} is knwon.

\begin{Def} {\bf (\cite{Ved})}
For mixed entangled states $\sigma$ on $\fH_1 \otimes \fH_2$, the relative entropy of entanglement is defined by
$$ E^R(\sigma) \equiv \min_{\rho \in \fD} U(\sigma \vert \rho),$$
where the minimum is taken for all $\rho \in \fD$, where $\fD$ represents the set of all separable (disentangled) states on $\fH_1 \otimes \fH_2$.
\end{Def}

This measure is a kind of distance (difference) between the entangled states $\sigma$ and the separable (disentangled) states $\rho$.
They also proposed the conditions that the entanglement-measure $E\left(\sigma\right)$ for any entangled states $\sigma$ on the total system $\fH_1 \otimes \fH_2$
 should satisfy. It is given in \cite{Ved} by 
\begin{itemize}
\item [(E1)] $E\left( \sigma  \right)=0\Leftrightarrow \sigma $ is separable.
\item [(E2)] $E\left( \sigma  \right)$ is invariant under the local unitary operations:
$$E\left( \sigma  \right)=E\left( {U_1\otimes U_2\sigma U_1^*\otimes U_2^*} \right), $$
where $U_i,(i=1,2)$ represent the unitary operators acting on $\fH_i$, $(i=1,2)$.
\item [(E3)] The measure of entanglement $E\left( \sigma  \right)$ can not be increased 
under the trace-preserving completely positive map given by $\Phi $. That is,
$$E(\Phi\sigma) \leq E(\sigma). $$
\end{itemize}

As a measure satisfying the above conditions, a special case of V.Vedral's definition, we introduced the entanglement degree due to the mutual entropy and then 
applied it to the analysis of the Jaynes-Cummings model in \cite{FO,FM,FN}. 
\begin{Def}  {\bf (\cite{FO,FM,FN})}
For mixed entangled states $\sigma$ on $\fH_1 \otimes \fH_2$, the entanglement degree due to the mutual entropy is given by
$$E^M(\sigma) \equiv Tr[\sigma (\log \sigma - \log \rho_1 \otimes \rho_2)], $$
where $\rho_1 \equiv tr_{\fH_2}\sigma$ and $\rho_2 \equiv tr_{\fH_1}\sigma$. 
\end{Def}
In the above definition, we fixed the separable (disentangled) state such as $\rho = tr_{\fH_2}\sigma \otimes tr_{\fH_1} \sigma$,
because it was difficult to find the separable (disentangled) state attaining the minimum value of the relative entropy of entanglement. 
The separable (disentangled) state chosen by $\rho = tr_{\fH_2}\sigma \otimes tr_{\fH_1} \sigma$
is nontrivial state but our measure contains both quantum and classical entanglement. That is, our measure takes greater value than V.Vedral's one. That is, from the
definitions, we easily find $E^R(\sigma) \leq E^M(\sigma)$.
 For example,
for pure entangled states, by the above Araki-Lieb's triangle inequality, we easily find that our measure is equal to the twice of von Nuemann entropy, namely 
$E^M(\sigma) = 2 E^R(\sigma)$ for pure entangled states $\sigma$, since $E^R(\sigma)$ becomes von Neuamm entropy for pure entangled states \cite{Ved2}.
However it was sufficient to get the rough degree of entanglement for the analysis of the time development of the Janeys-Cummings model. 

In this short paper, we adopt a parametrically extended entanglement-measure due to the Tsallis relative entropy which is a generalization of our previous entanglement-measure.
\begin{Def} {\bf (\cite{AR})}
For mixed entangled states $\sigma$ on $\fH_1 \otimes \fH_2$,  the entanglement-measure due to the Tsallis relative entropy is given by
$$E_q^T(\sigma) \equiv \frac{Tr[\sigma - \sigma ^q( \rho_1 \otimes \rho_2)^{1-q}]}{1-q}, $$
where $\rho_1 \equiv tr_{\fH_2}\sigma$ and $\rho_2 \equiv tr_{\fH_1}\sigma$. 
\end{Def}
Note that the above entanglement-measure is a special version of the generalized Kulback-Leibler measure of quantum entanglement introduced in \cite{AR}.
In addition, the above entanglement-measure for a non-trivial example was studied in \cite{AR}.
From the definition, we easily find that $\lim_{q\to 1} E_q^T(\sigma) = E^M(\sigma)$.

In the below, we show the equality condition of the inequality ((1) of Proposition \ref{prop1}) in the properties of the Tsallis relative entropy.
\begin{Lem}   \label{lemma_quantum_Tsallis}
For $q \in [0,1) \cup (1,2]$ and density operators $\rho$ and $\sigma$, we have
$$D_{q}(\rho\vert \sigma) \geq 0,$$
with equality if and only if $\rho = \sigma$.
\end{Lem}
{\it Proof}:
\begin{itemize}
\item[(1)] For $q \in (1,2]$ we have the lemma due to \cite{RS,FYK,OP}:
\begin{equation}   \label{quant_lemma_1}
D_{q}(\rho\vert\sigma) \geq Tr [\rho(\log\rho -\log\sigma)] \geq \frac{1}{2}Tr [\vert\rho-\sigma\vert]^2.
\end{equation}
\item[(2)] For $q \in [0,1)$ we also have
\begin{equation}  \label{quant_lemma_2}
D_{q}(\rho\vert\sigma) \geq Tr[\vert\rho-\sigma\vert].
\end{equation}
Indeed, for $x\geq 0,\,y\geq 0,\,$ and $0\leq q < 1$ we have 
$$ \frac{x-x^{q}y^{1-q}}{1-q} \geq x-y,$$
with equality if and only if $x=y$.
If we take spectral decompositions such that $\rho = \sum_i u_iP_i$ and $\sigma =\sum_jv_jQ_j$, then we have
\begin{eqnarray*}
D_{q}(\rho\vert\sigma) - Tr[\vert \rho -\sigma \vert] &=& Tr\left[ \frac{\rho-\rho^{q}\sigma^{1-q}}{1-q}-(\rho-\sigma) \right]\\
&=&\sum_{i,j} Tr \left[ P_i\left\{ \frac{\rho-\rho^{q}\sigma^{1-q}}{1-q}-(\rho-\sigma)\right\} Q_j \right]\\
&=&\sum_{i,j} Tr \left[ P_i\left\{ \frac{u_i-u_i^{q}v_j^{1-q}}{1-q}-(u_i-v_j)\right\} Q_j \right]\\
&=&\sum_{i,j} \left\{ \frac{u_i-u_i^{q}v_j^{1-q}}{1-q}-(u_i-v_j)\right\} Tr\left[ P_i Q_j \right] \geq 0.
\end{eqnarray*} 
\end{itemize}
Since $Tr[\vert X\vert ] =0$ implies $X=0$ \cite{Sch}, we have the lemma from Eq.(\ref{quant_lemma_1}) and Eq.(\ref{quant_lemma_2}).

\hfill \qed

\begin{Prop}
For $0 \leq q < 1$, $E_q^T$ satisfies the conditions (E1), (E2) and (E3) 
\end{Prop}
{\it Proof} : 
It is obvious from Proposition \ref{prop1} and Lemma \ref{lemma_quantum_Tsallis}.
\hfill \qed
\vspace*{0.5cm}

We also have the following proposition.

\begin{Prop}   \label{prop2}
\begin{itemize}
\item [(1)] For any entangled states $\sigma$ on $\fH_1 \otimes \fH_2$, we have $E_0^T(\sigma) = 0$.
\item [(2)] There exists $q$ in $[0,1)$ such that $E_q^T(\sigma) = E^R(\sigma)$ for any entangled states $\sigma$ on $\fH_1 \otimes \fH_2$.
\item [(3)] For any entangled states $\sigma$ and $\sigma'$ on $\fH_1 \otimes \fH_2$, and any $0 \leq q < 1$, we have the subadditivity:
$$E_q^T(\sigma \otimes \sigma ') \leq E_q^T(\sigma) + E_q^T(\sigma ').$$
\end{itemize}
\end{Prop}

{\it Proof} : 
(1) is trivial, since $\sigma^0 \equiv I$. 
We easily find from thier definitions that $0 \leq E^R(\sigma) \leq E^M(\sigma)$ as mentioned above, 
and $E_q^T(\sigma)$ continuously takes the values from $0$ to $E^M(\sigma)$. ($E_q^T(\sigma)$  is not necessarily monotone increase function for $q$.)
Therefore this assures that there exists $q$ in $0 \leq q <1$ such that $E_q^T(\sigma) = E^R(\sigma)$. 
Finally we show (3).
In general, we have the pseudoadditivity for the Tsallis relative entropy (see \cite{FYK} for example) :
$$ D_q(\rho_1\otimes\rho_2 \vert \sigma_1 \otimes \sigma_2) = D_q(\rho_1\vert \sigma_1)+D_q(\rho_2\vert \sigma_2)
 +(q-1)D_q(\rho_1\vert \sigma_1)D_q(\rho_2\vert \sigma_2) .$$
Thus we have 
\begin{equation}  \label{pseudo_E}
E_q^T(\sigma \otimes \sigma ') = E_q^T(\sigma) + E_q^T(\sigma ') + (q-1) E_q^T(\sigma)E_q^T(\sigma ').
\end{equation}
Since $E_q^T(\sigma)$ is nonnegative, we have the subadditivity $E_q^T(\sigma \otimes \sigma ') \leq E_q^T(\sigma) + E_q^T(\sigma ')$ for any $0 \leq q < 1$.
\hfill \qed
\vspace*{0.5cm}

We note that we have the additivity $E^M(\sigma \otimes \sigma ') = E^M(\sigma) + E^M(\sigma ')$ as $q \to 1$ in Eq.(\ref{pseudo_E}).
In addition,we shoud note that our measure $E_q^T(\sigma)$ takes $0$ when $q=0$, although $\sigma$ is not separable (disentangled) state.
%\end{Rem}
%\section{A few examples}  
Finally we give a simple example concerning on (2) of Proposition \ref{prop2}.
\begin{Ex}
We consider the famous Werner state which is a mixed entangled state :
$$W_F=F\left| {\Psi ^-} \right\rangle \left\langle {\Psi ^-} \right|+{{1-F} \over 3}\left( {\left| {\Psi ^+} \right\rangle \left\langle {\Psi ^+} \right|+\left| {\Phi ^-} \right\rangle \left\langle {\Phi ^-} \right|+\left| {\Phi ^+} \right\rangle \left\langle {\Phi ^+} \right|} \right)$$
where $\left| {\Psi ^\pm } \right\rangle ={1 \over {\sqrt 2}}\left( \left| \uparrow \right\rangle \otimes \left| \downarrow \right\rangle \pm \left| \downarrow \right\rangle \otimes \left| \uparrow \right\rangle \right),\left| {\Phi ^\pm } \right\rangle ={1 \over {\sqrt 2}}\left( \left| \uparrow \right\rangle \otimes\left| \uparrow \right\rangle \pm \left| \downarrow \right\rangle \otimes \left| \downarrow \right\rangle \right)$.

Then the relative entropy of entanglement was calculated \cite{Ved,Ved2}
\[
E^R \left( {W_F } \right) = \left\{ \begin{array}{l}
 \,\,\,\,\,\,\,\,\,\,\,\,\,\,\,\,\,\,\,\,\,\,\,\,\,\,\,\,\,\,\,\,\,\,0\,\,\,\,\,\,\,\,\,\,\,\,\,\,\,\,\,\,\,\,\,\,\,\,\,\,\,\,\,\,\,\,\,\,\,\,\,\,\,\,\,\,\,\,\,\,\,\,\,\,\,\,\,\,\,\,\,\,\,\,\,\,\,\,\,\,\,\,\,\,\,\,\,\,\,\left( {F \le 0.5} \right) \\ 
 F\log F + \left( {1 - F} \right)\log \left( {1 - F} \right) + \log 2\,\,\,\,\,\,\,\,\,\,\,\,\,\,\,\,\,\left( {F \ge 0.5} \right). \\ 
 \end{array} \right.
\]
On the other hand, our measure $E_q^T(W_F)$ is calculated by
$$
E_q^T(W_F) = \frac{1-(\frac{1}{4})^{1-q}F^q+(\frac{3}{4})^{1-q}(1-F)^q}{1-q}.
$$

Figure 1 shows that when $q=0.35$, $E_{q}^T\left(W_F\right) \simeq  E^R \left( {W_F } \right)$ for $F \simeq  0.9$ and   $F \simeq  0.98$. 
For different values of the parameter $F$, we can find the parameter $q$ in $0\leq q <1$ satisfying $E_q^T(W_F) = E^R(W_F)$ thanks to (2) of Proposition \ref{prop2}.

\begin{figure}[htbp]
\begin{center}
\includegraphics{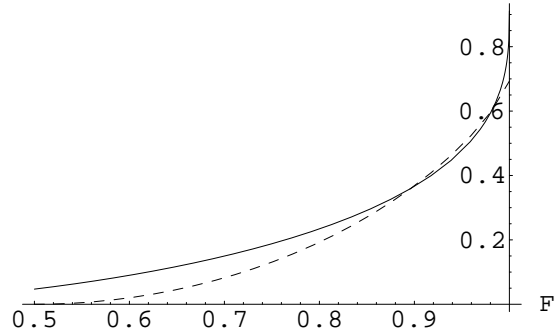}
\caption{Entanglement degree of Werner state $W_F$ by $E_{0.35}^T\left(W_F\right)$ (solid line) and $E^R \left( {W_F } \right)$ (dashed line).}
\end{center}
\end{figure}
\end{Ex}

%\begin{Ex}
%For
%\[
%\sigma _1  = \lambda \left| {\Phi ^ +  } \right\rangle \left\langle {\Phi ^ +  } \right| 
%+ \left( {1 - \lambda } \right)\left| {\uparrow  \otimes \downarrow } \right\rangle \left\langle {\uparrow  \otimes \downarrow } \right| .
%\]
%$E^R(\sigma_1)$ was calculated \cite{Ved2}
%\[
%E^R \left( {\sigma _1 } \right) = \left( {\lambda  - 2} \right)\log \left( {1 - \frac{\lambda }{2}} \right) + \left( {1 - \lambda } \right)\log \left( {1 - \lambda } \right).
%\]
%Our measure $E_q^T(\sigma_1)$ is calculated by
%$$
%E_q^T(\sigma_1) = \frac{1-4^{q-1}\left\{(1-\lambda)^q (\lambda -2)^{2(1-q)}+\lambda^q(\lambda (2-\lambda) )^{1-q}  \right\}}{1-q}.
%$$
%\end{Ex}

%\begin{Ex}
%For
% \[
%\sigma _2  = \lambda \left| {\Phi ^ +  } \right\rangle \left\langle {\Phi ^ +  } \right| + \left( {1 - \lambda } \right)\left| {0 \otimes 0} \right\rangle \left\langle {0 \otimes 0} \right|
%\]
%$E^R(\sigma_2)$ was calculated \cite{Ved2}
%\[
%E^R \left( {\sigma _2 } \right) = s_ +  \log s_ +   + s_ -  \log s_ -   - \left( {1 - \frac{\lambda }{2}} \right)\log \left( {1 - \frac{\lambda }{2}} \right) - \left( {1 - \frac{\lambda }{2}} \right)\log \left( {1 - \frac{\lambda }{2}} \right).
%\]
%\end{Ex}

\end{document}